\documentclass[reprint, secnumarabic, amssymb, nobibnotes, aps, prl, floats, floatfix]{revtex4-1}

\usepackage{bm}

\usepackage{graphicx}
\usepackage[export]{adjustbox}
\usepackage{amsmath}
\usepackage[dvipsnames]{xcolor}
\usepackage{empheq}

\usepackage{changes}
\usepackage[colorinlistoftodos, prependcaption]{todonotes}
\usepackage{xargs}    
\newcommandx{\improvement}[2][1=]{\todo[linecolor=Plum,backgroundcolor=Plum!25,bordercolor=Plum,#1]{#2}}
\newcommandx{\change}[1]{\todo[color=blue!20!green!20,  bordercolor=white, inline]{#1}}
\newcommandx{\old}[1]{}
\newcommandx{\new}[1]{#1}

\begin{document}

\author{Tobias Pfeffer}
\affiliation{Department of Physics, Arnold Sommerfeld Center for Theoretical Physics,
University of Munich, Theresienstrasse 37, 80333 Munich, Germany}

\author{Zhiyuan Yao}
\affiliation{Department of Physics, University of Massachusetts, Amherst, MA 01003, USA}

\author {Lode Pollet}
\affiliation{Department of Physics, Arnold Sommerfeld Center for Theoretical Physics,
University of Munich, Theresienstrasse 37, 80333 Munich, Germany }
\title{Strong randomness criticality in the scratched-XY model}%


\date{\today}%

\begin{abstract}  
We study the finite-temperature superfluid transition in a modified two-dimensional (2D) XY model with power-law distributed ``scratch"-like bond disorder. 
As its exponent decreases, the disorder grows stronger and the mechanism driving the superfluid transition changes from conventional vortex-pair unbinding to a strong randomness criticality (termed scratched-XY criticality) characterized by a non-universal jump of the superfluid stiffness.
The existence of the scratched-XY criticality at finite temperature and its description by an asymptotically exact semi-renormalization group theory, previously developed for the superfluid-insulator transition in one-dimensional disordered quantum systems, is numerically proven by designing a model with minimal finite size effects.
Possible experimental implementations are discussed.
\end{abstract}

\maketitle

It is well known that in spatial dimensions $\text{D} \leq 2$ long-range order is destroyed by thermal fluctuations for systems with continuous symmetry and short-range interactions \cite{MerminWagner}.
However, the 2D XY-model describing the superfluid to normal liquid (SF--NL) transition at finite temperature can still undergo a Berezinskii-Kosterlitz-Thouless (BKT) transition driven by the proliferation of topological defects, in particular the unbinding of vortex and anti-vortex pairs \cite{KT, BKT, KTRG}. 
This transition features a universal jump of the superfluid stiffness $\Lambda$ at the critical temperature $T_{c}$, \emph{ i.e.}, $\Lambda(T_{c}) /  T_{c}= 2 / \pi$ at the transition.\\
The question whether there exists an alternative mechanism for the destruction of superfluidity fundamentally different from the proliferation of topological defects has been a contentious one for several decades -- especially in the context of the one-dimensional (1D) superfluid--Bose-glass (BG) quantum phase transition \cite{GSI, GSII, GShighloop}. 
While in the weak disorder regime the transition is driven by the proliferation of instanton--anti-instanton pairs [``vertical'' vortex--anti-vortex pairs in the $(1+1)$-dimensional superfluid phase field] with a universal critical Luttinger liquid parameter $K_{c} = 3/2$ \cite{GSpopov}, the possibility of a different mechanism in the strong disorder regime can not be ruled out. 
Using the strong disorder renormalization group (SDRG) method, Altman \emph{et al.} claimed that the Coulomb blockade physics of weak links (strong potential barriers) can give rise to a new criticality in the strong disorder regime \cite{realspaceRGI, realspaceRGII, realspaceRGIII, realspaceRGIV}. 
However, in this case the SDRG is uncontrolled as the fixed point solutions violate the assumptions under which the approximate RG equations have been derived.
Based on the Kane-Fisher physics of weak links \cite{KaneFisher_short, KaneFisher_long}, Pollet \emph{et al.} developed an asymptotically exact theory of the 1D superfluid-insulator transition and showed that rare weak links can destroy superfluidity and give rise to a new criticality, the so-called scratched-XY (sXY) criticality \cite{sXYI,sXYII,sXYIII}.
The hallmark of the transition is the relation $K_c =  1 /\zeta$, where $\zeta$ is a microscopic, irrenormalizable parameter characterizing the scaling behavior of the bare strength of the typically weakest links, $J_{0}^{(L)} \sim 1 / L^{1-\zeta}$, in a system of size $L$.\\
However, the explicit relationship between $\zeta$ and the microscopic parameters is unknown, and extracting $\zeta$ numerically or experimentally requires great effort.
Strong finite size effects in the 1D Bose-Hubbard model with diagonal disorder where so far preventing a solid numerical proof for the validity of the sXY scenario \cite{sXYIII} -- 
even to the extent that despite several large-scale simulations, a consensus of the nature of the superfluid-insulator transition in the strong disorder regime has not been reached \cite{Gerster, Vojta, realspaceRGIV, Doggen}.\\
In this Letter, we study the superfluid transition in a classical XY model with power-law distribution of parallel ``scratches". Due to the simplicity of this model $\zeta$ can be determined analytically. 
These properties enable us to unambiguously demonstrate the existence of the sXY university class and verify the theory by Pollet \emph{et al.} for 1D superfluid-insulator transitions.
We also show that the theory by Altman \emph{et al.} fails to describe the strong disorder critical point.  
Moreover, thanks to the fact that in the scratched-XY model $\zeta$ is controlled by a microscopic parameter an experimental verification for this new criticality for a finite temperature phase transition is feasible.
\\
\\
{\it The scratched-XY model} --
The Hamiltonian of our scratched-XY model reads
\begin{equation} \label{XYmodel}
	H = -\sum_{\bm{r},\hat{\mu}} J_0(\bm{r},\hat{\mu}) \cos(\theta_{\bm{r}} - \theta_{\bm{r}+\hat{\mu}}),
\end{equation}
where $\bm{r} = (x, y)$ is the site index of the square lattice,  $\hat{\mu} \in \{ \hat{x}, \hat{y} \}$ a unit vector along the bonds, and $J_{0}(\bm{r}, \hat{\mu})$ the corresponding coupling. 
Our units are $J_{0}(\bm{r}, \hat{y}) = 1$ and lattice spacing $a=1$.
The probability distribution of $J_{0}(x, \hat{x})$ is taken to be a power law distribution,
\begin{equation} \label{disorderdistribution}
    p(J_0) \text{d}J_0 = \frac{1}{\Gamma} J_0^{1/\Gamma - 1} \text{d}J_0\, , \qquad J_{0}  \in [0, 1] \, ,
\end{equation}
where $\Gamma < 1$ is the only parameter of the model. From the following discussion it will become clear that $T_c(\Gamma = 1)=0$ in analogy with the diluted Ising model \cite{SachdevQPT}. 
The bare strength of the typically deepest scratch $J_{0}^{(L)}$ in a square lattice with linear size $L$ can be estimated by imposing that finding at least one such deep scratch has a probability of order one, 
\begin{equation} \label{probscratch}
   L  \int_{0}^{J_{0}^{(L)}} \!\!\!\! p(J_{0}) d J_{0} \sim 1 \, .
\end{equation}
Therefore, $J_{0}^{(L)}$ scales with $L$ as a power law, 
\begin{equation} \label{eq:Jscaling}
    J_{0}^{(L)} \sim \frac{1}{L^{1 - \zeta}}  \quad \text{where } \zeta = 1 - \Gamma \, .
\end{equation}
Another property of the distribution is that on every new length scale the expectation value of the number of the typically deepest scratches corresponding to the new scale is just one. This follows directly from (\ref{probscratch}).
Because of the presence of deep scratches with $J_{0} \ll 1$, starting from mesoscopic scales, the system can be viewed as superfluid regions joined by barriers formed by single or consecutive scratches.
Therefore, in addition to the topological defects, the superfluid stiffness in the $x$-direction will be renormalized by the barriers connecting adjacent superfluid regions.
Quantitatively, the action $S$ that describes an otherwise homogeneous superfluid system with a barrier at $x=0$ is
\begin{equation} \label{effectiveaction}
    S = \sum_{i} \frac{K}{2\pi} \int \!\! d x dy \, (\nabla \theta_{i})^{2}  - \frac{t}{T} \int \!\! dy \, \cos (\theta_{+} - \theta_{-}) \, .
\end{equation}
We have rescaled $x$ and $y$ and introduced a dimensionless number $K = \pi \sqrt{\Lambda_{x} \Lambda_{y}} / T$ with $\Lambda_{x}, \Lambda_{y}$ the superfluid stiffness in $x , y$ direction respectively. 
Here $\theta_{i} \, (i=1,2)$ is the phase field of the left and right superfluid, $\theta_{-}$ and $\theta_{+}$ are the values of the left and right phase field at $x=0$, and $t$ is proportional to the bare strength of the barrier.
The renormalization of the strength of the barrier by harmonic modes in the phase field is described by the Kane-Fisher flow equation \cite{KaneFisher_short, KaneFisher_long}
\begin{equation} \label{KaneFisherRG}
    \frac{\text{d} t(\ell) }{ \text{d} \ell } = \left( 1 - K^{-1} \right) t(\ell) \, ,
\end{equation} 
where $t(\ell)$ is the renormalized strength of the barrier at length scale $\ell = \ln L$.
Since the critical value $K_{c} \geq 2$ (vortex--anti-vortex pairs will proliferate below $K=2$), the bare strength $t$ will be renormalized towards strong couplings and the RG flow (\ref{KaneFisherRG}) stops at the \emph{clutch scale} $\ell^{*}$ where $t(\ell^{*}) / T \sim 1$ \cite{sXYII, sXYIII}. 
When the clutch scale is reached, the system size has been rescaled by a factor $ 1/ L^{*}$ with $L^{*} = \exp(\ell^{*})$. At scales much bigger than the clutch scale, the effect of the barrier on renormalizing the superfluid stiffness is
\begin{equation}  \label{eq: JL}
    \Lambda^{-1}_{x}(\ell) - \Lambda^{-1}_{x}(\ell_{0}) \propto \frac{1}{t(\ell^{*}) L / L^{*}} \propto   \frac{L^{*}}{L} \, ,
\end{equation}
where $\ell_{0}$ is some mesoscopic scale. \\
In the following we assume that the barriers are formed by single scratches and consecutive scratches play a subdominant role (numerically justified later).
In the scratched-XY model with well-separated typical deepest scratches, cf. Eq.~(\ref{probscratch}), it is possible to write down a flow equation which accounts for the renormalization effect of the scratches on different length scales successively. Moreover, the theorem of self-averaging \cite{sXYI} allows us to write the RG equation in terms of the median of $\Lambda^{-1}_{x}$ instead of the full distribution. This theorem guarantees that the distribution of the superfluid stiffness (along the $x$ direction in our case) flows towards a $\delta$-like distribution in the superfluid phase including the critical point \cite{sXYI}.
The flow of $\Lambda^{-1}_{x}$ (in the median sense) due to the scratches is given by
\begin{equation} \label{stiffnessRG}
    \frac{\text{d} \, \Lambda_{x}^{-1}(\ell)}{\text{d} \, \ell }  \propto  \frac{L^{*}}{L} \, .
\end{equation}
\begin{figure}[t]
\includegraphics[width = 0.45 \textwidth]{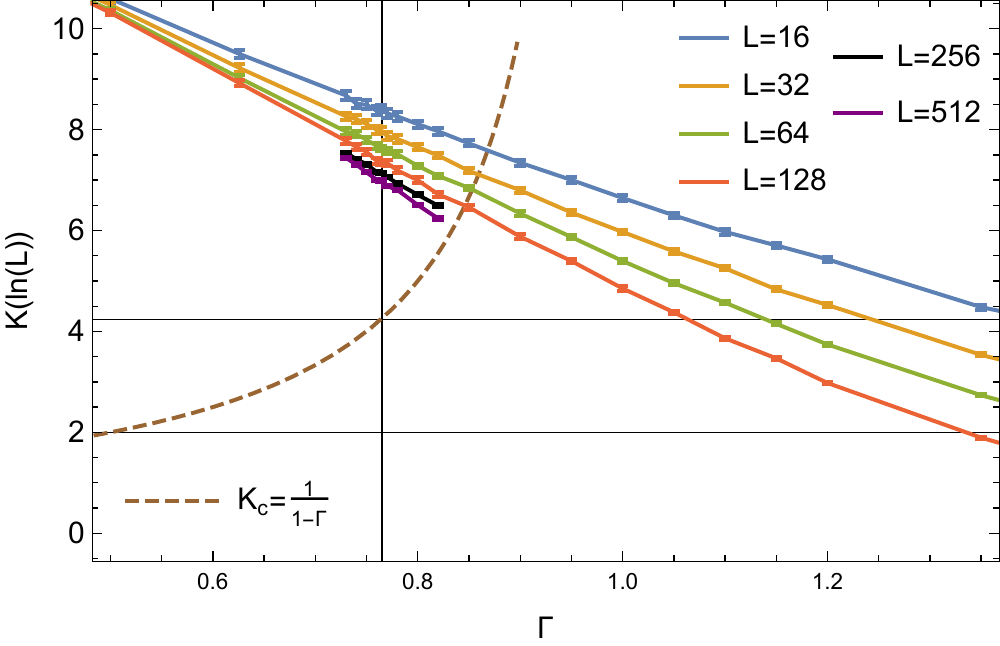}
\caption{ (color online) A plot of the disorder averaged $K( \ln L)$ for $L=16,  32, \dots, 512$. The brown dashed line is a plot of the critical $1/\zeta$ line. We find a critical value of  $\Gamma_c = 0.764(2)$ (vertical grid line) with a non-universal value of $K_c = 4.24(4)$ (upper horizontal grid line) at the transition. The non-universal value of $K_c$ is larger than  in the BKT case where $K_{c} =2$ (lower horizontal grid line). 
}
\label{fig:BareDataK}
\end{figure}
\begin{figure}[t]
\includegraphics[width = 0.45 \textwidth]{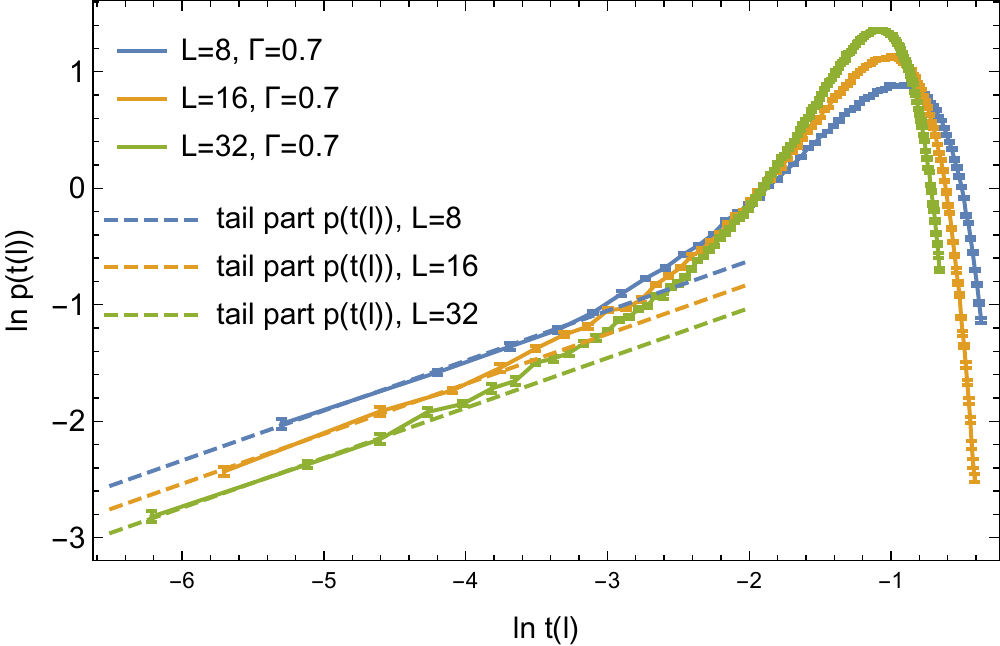}
\caption{
(color online) The log-log plot of the distribution of renormalized strengths $t(\ell)$ at $\Gamma=0.7$ for $L = 8, 16, 32$ from $8 \times 10^{5}, 1.2 \times 10^{6}, 2.0 \times 10^{6}$ disorder realizations (in contrast to a few thousands in \cite{Doggen}).
The slope of the tail keeps changing until small enough $t(\ell)$ (requiring a sufficient number of disorder realizations) is reached. The tail part on the log-log scale is perfectly fitted by a linear line with a slope $0.43(1)$ which agrees with the exponent $1/\Gamma -1$ within error bars. Therefore, the barriers joining adjacent superfluid regions are formed by the individual deepest scratches.}
\label{fig:zeta}
\end{figure}
Rewriting Eq.~(\ref{stiffnessRG}) in terms of the parameter $K$ and introducing  $w(\ell) = L^{*} / L$ leads to
\begin{equation} \label{eq: RG_K}
    \frac{\text{d} K(\ell )}{\text{d} \ell } = - w \; K^3 \, ,
\end{equation}
where we have rescaled $w$ to absorb unimportant coefficients.
The clutch scale implicitly depends on the system size through the typical deepest scratch, cf.~(\ref{eq:Jscaling}).
Together with (\ref{KaneFisherRG}), the RG equation for $w(\ell)$ reads \cite{sXYII, sXYIII}
\begin{equation} \label{eq: RG_w}
    \frac{\text{d} w(\ell)}{\text{d} \ell } = \frac{1 - \zeta K}{K-1} w .
\end{equation}
Therefore, for $1/\zeta_c > 2$, a new strong randomness criticality emerges where the superfluid transition is driven by scratches and the vortex--anti-vortex pairs play a subdominant role.
Consequently, we can neglect the vortex--anti-vortex pairs in studying this new criticality, and the critical condition is given by 
\begin{equation} \label{eq:K_zeta_relation}
	K_{c} = 1/\zeta_{c}.
\end{equation}
\\
{\it RG flow} --
Near the strong randomness critical point, it is convenient to introduce $x(\ell) = K(\ell) - \zeta_c^{-1}$ and linearize the RG equations (\ref{eq: RG_K}), (\ref{eq: RG_w}), 
\begin{subequations} \label{eq:RG_linear}
    \begin{empheq}{align}
        \frac{\text{d} \tilde{x}}{\text{d} \ell} &= - \tilde{w} \\
        \frac{\text{d} \tilde{w}}{\text{d} \ell} &= -2 \tilde{x} \tilde{w} \, ,
    \end{empheq}
\end{subequations}
where $\tilde{x} = x \; \zeta_{c}^{2} (1-\zeta_{c})^{-1}/2$ and $\tilde{w} = w /\,  (2\, \zeta_c (1-\zeta_{c}))$ are rescaled $x$ and $w$, respectively.
The RG invariant $A = \tilde{w} - \tilde{x}^{2}$ is an analytic function of the microscopic parameters $\zeta$ and $T$, and $A=0$ corresponds to the critical flow. At fixed temperature and near the critical point $(T, \zeta_{c})$, $A \approx B (\zeta_{c} - \zeta)$ where $B$ is a constant, and $\zeta = 1 - \Gamma$ acts as the tuning parameter.  
The solution $\tilde{x}(l)$ away from the critical point is given by
\begin{equation}
    \tilde{x}(\ell) =  \frac{\sqrt{|{A}|}}{f \left( \sqrt{|{A}|} ( \ell  + C) \right)} \, ,
\label{noncriticalFlow}
\end{equation}
where $A > 0$ , $f(z) = \tanh z$ on the superfluid side, $A < 0$, $f(z) = \tan z$ on the disordered side, and $C$ is another RG invariant.
The flow at the critical point ($A=0$) is given by
\begin{equation} \label{criticalFlow}
    \tilde{x}(\ell) =  \frac{1}{\ell + C} \, .
\end{equation}
The solutions of the RG equations (\ref{eq:RG_linear}) are used to extrapolate finite size data to infinite system size. To this end, we define the universal scaling function $F(z)$, 
\begin{equation}
     F(z) \equiv (\ln L + C)   \left[ K(\zeta, \ln L)  - \zeta_{c}^{-1} \right]
\label{scalingForm}
\end{equation}
where $z = (\zeta_{c} - \zeta) (\ln L + C)^{2}$.
The universal scaling function $F(z)$ has the property $F(0) = 2(1-\zeta_c) / \zeta^2_c$.\\
\\
{\it Numerical simulation} --
\begin{figure}[t]
\includegraphics[width = 0.45 \textwidth]{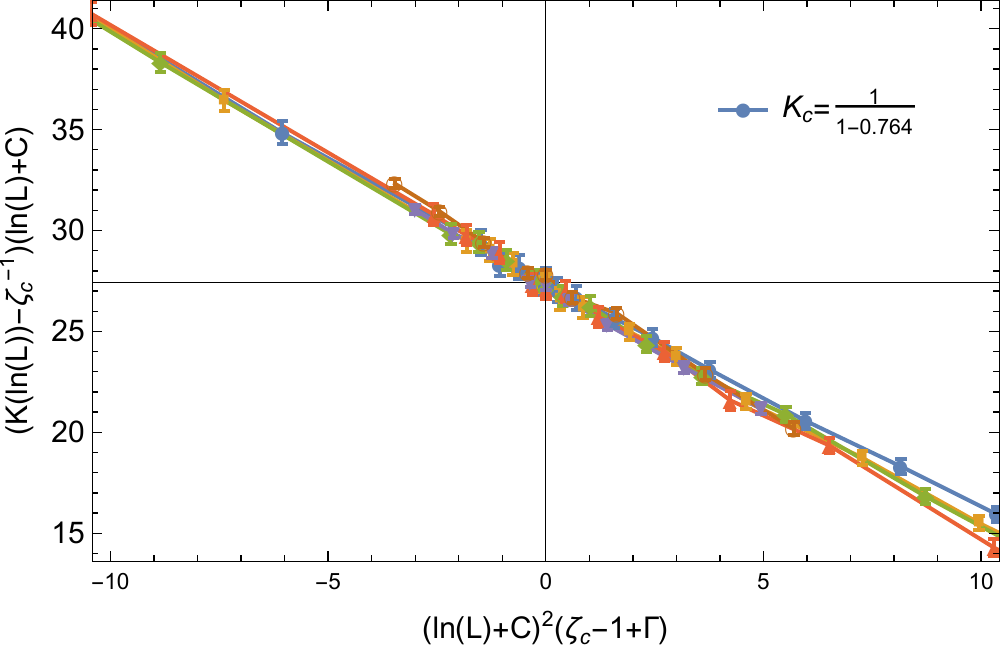}
\caption{
(color online)  Shown is the data collapse of $\left( K(\ell) - \zeta_c^{-1} \right) \left( \ln( L) + C \right)$ over $(\zeta_{c} - \zeta)(\ln L + C)^{2}$. With $\zeta_{c} = 0.236(4)$ and $C = 3.86(5)$, all the finite size data collapse onto a single line satisfying the constraint $F(0) = 2(1-\zeta_{c})/\zeta_{c}^{2}$. The critical $K_{c}$ is given by $K_{c} = 1/\zeta_{c} = 4.24(7) > 2$ as predicted by the strong randomness criticality.
}
\label{fig:CollapseDataK}
\end{figure}
To numerically establish the strong randomness criticality, we study the superfluid response of the scratched-XY model at fixed temperature, $T = 0.2$, by tuning $\Gamma$.
For a square lattice with linear size $L$, we first draw $L$ random scratches $J_0$ according to the power law distribution (\ref{disorderdistribution}).
We then perform simulations by using the classical Worm algorithm \cite{classicalWorm}.
In writing down Eq.~(\ref{stiffnessRG}), we assumed that the barriers joining adjacent superfluid regions are formed by the single deepest scratches, \emph{i.e.} the distribution of the bare strength of the barriers $p(t)$ at large length scales is given by (\ref{disorderdistribution}).
We justify this assumption by studying the distribution of the renormalized barrier strengths, $p(t(\ell))$, from a large number of disorder realizations in systems with mesoscopic system sizes $L$.
Since strong barriers act as Josephson junctions, the supercurrent response $j$ under a phase twist $\varphi$ is given by
\begin{equation}
    j = \frac{\partial F}{\partial \varphi} = t(\ell) \exp( - TL/ 2 \Lambda)  \sin \varphi,
\end{equation}
where $F$ is the free energy of the system under a phase twist in the $x$-direction, $\Lambda$ is the superfluid stiffness of the left and right superfluids, $t(\ell)$ is the renormalized strength of the barrier, and $\exp( - TL/ 2 \Lambda )$ accounts for the effect of supercurrent states at finite temperature \cite{NB2000}.
The renormalized strength $t(\ell)$ can then be readily related to the winding number fluctuations in the $x$-direction by taking a second order derivative of $F$ with respect to $\varphi$.
Since $t(\ell)$ is determined through the supercurrent response under a phase twist across the system, $t(\ell)$ is determined irrespective of the microscopic origins.
Since the clutch scales of anomalously strong barriers will be much bigger than $L$, they will pick up a common factor due to the Kane-Fisher renormalization (\ref{KaneFisherRG}). 
Consequently, the tail of the distribution of the renormalized barrier strengths $p(t(\ell))$ will be the same as the distribution of the bare barrier strengths.
As can be seen from Fig.~\ref{fig:zeta}, the tail part of the distribution of $p(t(\ell))$ is described by the same power-law distribution as (\ref{disorderdistribution}).
Therefore, the barriers joining adjacent superfluid regions are formed by the single deepest scratches.
Moreover, the power law exponent of the tail of $p(t(\ell))$ does not flow with system size.
This is in sharp contrast to the theory of Altman \emph{et al.} which predicts a flow of the power law exponent governing the tail of the distribution of the renormalized strength of the barriers, \emph{i.e.} strong barriers are joined to form even stronger barriers.
The value of $t(\ell)$ at which this power law behaviour sets in decreases for increasing system sizes.
Therefore, a large number ($>10^{6}$) of disorder realizations is needed to resolve the genuine tail behavior.\\
Having justified the key assumption in deriving the strong randomness RG equations, we continue to perform measurements for different system sizes in order to verify the sXY criticality.
The superfluid stiffness is related to the winding number statistics by the Pollock-Ceperley formula \cite{Ceperley}, 
\begin{equation}
    \Lambda_{\mu} = T \langle W_{\mu}^2 \rangle \, ,
\end{equation}
where $\mu \in \{x, y\}$ is the label of spatial direction, $W_{\mu}$ is the winding number in that direction, and $\langle \cdots \rangle$ refers to statistical averaging. 
Since the RG equations (\ref{eq: RG_K}), (\ref{eq: RG_w}) can also be understood in terms of the mean of the full distribution (medians are only needed for the inverse quantities) we average over a big number of disorder realizations (typically 5000 or more) to extract the observables. 
To determine the strong randomness critical point, we need to extract $K(\infty)$ from our finite size data. This is accomplished by the previously discussed data collapse technique. With the choice of $\zeta_{c} = 0.236(4), C = 3.86(5)$, all the finite size data fall onto a single line within error bars. From Eq.~(\ref{eq:K_zeta_relation}), the critical value of $K$ is $K_{c} = 1/\zeta_{c} = 4.24(7)$ consistent with the condition of the strong random criticality $K_{c} > 2$.
\begin{figure}[t]
\includegraphics[width = 0.45 \textwidth]{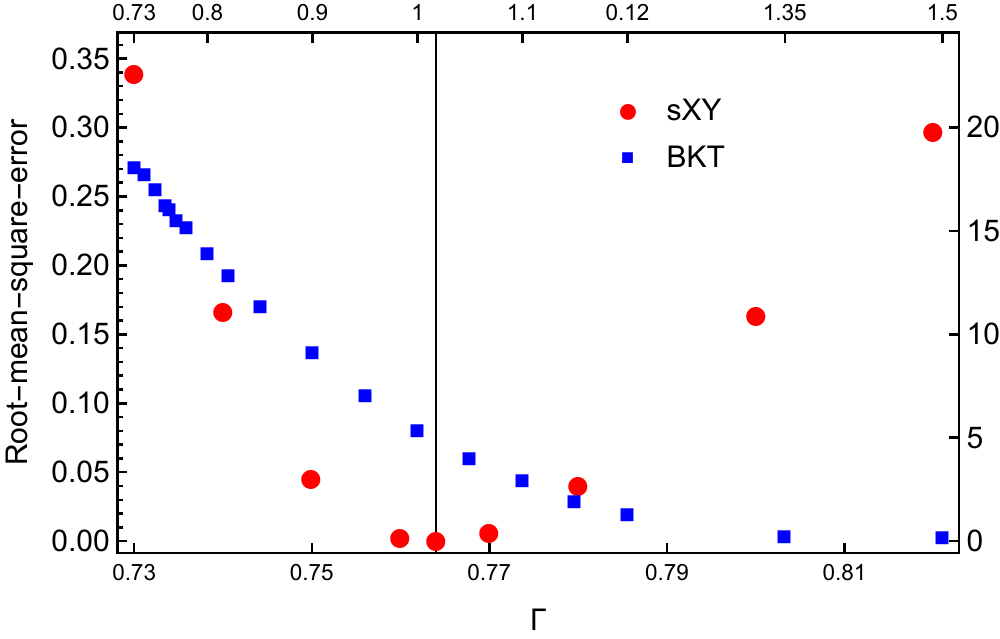}
\caption{
(color online) The Weber-Minnhagen \cite{WeberMinnhagen} root-mean-square-error $\sigma$ by fitting the flow of $K(\ln L)$ for $L= 16, 32, 64, 128, 256, 512$ to the critical flow of the sXY criticality (red dots, lower and left axis) and to the BKT criticality (blue squares, upper and right axis). While $\sigma$ displays a sharp minimum at  $\Gamma_c = 0.764(2)$ for the sXY RG, there is no such minimum for the BKT RG. 
}
\label{fig:WeberMinnhagenFit}
\end{figure}
That the numerically obtained flow of $K$ is described by the sXY scenario is further supported by performing a single parameter Weber-Minnhagen fit of our finite size data to the critical RG flow for different values of $\Gamma$. For the flow at the critical point, the root-mean-square-error $\sigma$ is expected to show a sharp minimum \cite{WeberMinnhagen}. As shown in Fig.~\ref{fig:WeberMinnhagenFit}, $\sigma$ indeed exhibits a sharp minimum at a point, \emph{i.e.}, the critical point.  For completeness, we also demonstrate that $\sigma$ does not display a sharp minimum for a fit to the critical BKT flow in a broad region where the phase transition, if any, should occur. The BKT critical flow is given by 
\begin{equation}
    K(\ell) = 2 + \frac{1}{\ell + C} \, ,
\label{criticalFlowBKT}
\end{equation}
where $K_{c}=2$ from the Nelson-Kosterlitz relation \cite{NelsonKosterlitz}.
As can be seen from Fig.~\ref{fig:BareDataK}, at $\Gamma = 1.35$, $K(\ln 128)$ is already smaller than the universal value $2$. 
Therefore, $\Gamma_{c,\text{BKT}} < 1.35$ as $K(\ln L)$ decreases monotonically along the RG flow. However, no minimum for $\sigma$ can be found, cf. Fig.~\ref{fig:WeberMinnhagenFit}, implying that the RG flow can not be captured by the BKT criticality.
\\
\\
{\it Conclusion and Outlook} --
In summary, we have established that superfluidity in a 2D XY model with disordered scratches can be destroyed by a mechanism fundamentally different than the proliferation of vortex--anti-vortex pairs. 
The Kane-Fisher physics of scale-dependent scratches provides an alternative mechanism for destroying superfluidity in the strong disorder regime.
A key feature of the RG equations describing this new criticality is that a microscopic, irrenormalizable parameter $\zeta$ enters the equations and determines the non-universal jump of the superfluid stiffness at the transition point.
We introduced a minimal model in which $\zeta$ was readily related to the power law exponent $\Gamma$ characterizing the disorder distribution of the scratches.
At $T = 0.2$, we have determined $\Gamma_c = 0.764(2)$ and $K_{c} = 4.24(4)$, consistent with the strong disorder scenario $K_c > 2$.
Our analysis and simulations rule out all the other scenarios presented for the superfluid transition in the strong disorder regime.
The scratched-XY model can be realized in 2D Josephson junction arrays where the individual phase fields of the superconducting islands can establish global phase coherence due to the tunneling of Cooper pairs between the islands \cite{JosephsonEffect}, \emph{i.e.}, disorder can directly couple to the phase field through the strength of the tunneling barrier. Existing techniques make it possible to study the BKT-transition in 2D Josephson junction arrays \cite{JJA}. In order to introduce disorder in this systems such that the power law exponent of its distribution can be determined the strength of the tunneling barriers have to be controlled to high accuracy.

\begin{acknowledgements}
{\it Acknowledgement} --
We thank N. V. Prokof'ev, B. V. Svistunov and A. Kuklov for enlightening discussions. TP and ZY acknowledge the hospitality of the Flatiron Institute, New York City.
This work was supported by H2020/ERC Consolidator Grant No. 771891 (QSIMCORR), the Munich Quantum Center and the DFG through Nano-Initiative Munich, the Simons Collaboration on the Many Electron Problem, and the National Science Foundation under grant DMR-1720465.
The open data for this project can be found at \url{https://gitlab.lrz.de/QSIMCORR/scratchedXY}.
\end{acknowledgements}
\end{document}